\documentclass[conference]{IEEEtran}
\IEEEoverridecommandlockouts
\usepackage{cite}
\usepackage{amsmath,amssymb,amsfonts}
\usepackage{algorithmic}
\usepackage{graphicx}
\usepackage{booktabs}
\usepackage{textcomp}
\usepackage{multirow}
\usepackage{xcolor}
\def\BibTeX{{\rm B\kern-.05em{\sc i\kern-.025em b}\kern-.08em
    T\kern-.1667em\lower.7ex\hbox{E}\kern-.125emX}}

\begin{document}

\title{Sequential and Generative Models for Vehicular Distributed MIMO Channel Prediction\\

}

\author{
\IEEEauthorblockN{Malek ABIDA}
\IEEEauthorblockA{\textit{College of computing} \\
\textit{GVSU}\\
Michigan, USA \\
abidam@mail.gvsu.edu}
\and
\IEEEauthorblockN{Rafik ZAYANI}
\IEEEauthorblockA{\textit{Univ Rennes, CNRS, IETR,}\\
\textit{UMR 6164, F-35000}\\
Rennes, France \\
rafik.zayani@univ-rennes.fr}

\and
\IEEEauthorblockN{Eric P. SIMON}
\IEEEauthorblockA{\textit{Univ Lille, CNRS, IEMN,}\\
\textit{UMR 8520, F-59650}\\
Lille, France \\
eric.simon@univ-lille.fr}
\and
\IEEEauthorblockN{Davy P. GAILLOT}
\IEEEauthorblockA{\textit{Univ Lille, CNRS, IEMN,}\\
\textit{UMR 8520, F-59650}\\
Lille, France \\
davy.gaillot@univ-lille.fr}

}

\maketitle

\begin{abstract}
Vehicular communication is a key 6G use case requiring reliable and high-capacity connectivity under fast mobility and highly time-varying propagation conditions. However, large-scale vehicular channel estimation is costly and limited, impacting system-level performance of vehicular communications, and realistic channel prediction models are needed. This paper proposes a vehicular channel prediction framework based on real measured urban channels collected through a dedicated measurement campaign using the MaMIMOSA channel sounder. The framework enables the training and systematic benchmarking of sequential and generative models for both single-step and multi-horizon vehicular channel state information (CSI) prediction to assess prediction robustness across different forecasting horizons, including LSTM, TCN, a CNN-enhanced Transformer, and ChannelGPT, with the goal of accurately predicting channel evolution while preserving spatiotemporal dynamics and non-stationarity. In addition, a system-level evaluation framework is introduced to assess the impact of channel prediction on the performance of vehicular distributed MIMO communications. Using predicted channels, spectral efficiency (SE) is evaluated against true CSI. Results show that ChannelGPT achieves over 94\% normalized mean squared error (NMSE) reduction compared to LSTM and significant improvements over other baselines, while reducing FLOPs by 28\% and inference latency by 39\% relative to the CNN + Transformer. Moreover, ChannelGPT-predicted channels yield SE distributions nearly indistinguishable from those obtained with real measurements, demonstrating its effectiveness for reliable performance evaluation in high-mobility 6G vehicular networks.
\end{abstract}


\section{Introduction}

Vehicular communication is a cornerstone of emerging 6G networks, enabling ultra-reliable, low-latency, and high-capacity connectivity for safety-critical and high-mobility applications \cite{saeed_6g_vnet}. Achieving these performance targets critically depends on accurate channel estimation and prediction under rapidly time-varying propagation conditions, where channel estimation acquires current channel state information (CSI) through pilot-based measurements, whereas channel prediction forecasts future CSI from previously observed channel realizations. In this work, the considered learning models operate as channel predictors and are evaluated according to their ability to predict future vehicular CSI. However, acquiring large-scale vehicular channel datasets from real-world measurements remains costly and time-consuming. Measurement campaigns are constrained by hardware availability, bandwidth, and operational complexity, while classical analytical channel models often fail to capture the strong non-stationarity, and mobility-induced dynamics observed in realistic vehicular environments \cite{survey_v2x,muhammad_6g_v2x}. As a result, learning-based methods trained solely on idealized or simulated data exhibit limited generalization in practice.
Recent advances in deep learning have shown strong potential for enhancing channel modeling and prediction. Nevertheless, modern data-driven models are inherently data-hungry and require diverse, representative datasets to generalize reliably \cite{deepmimo}. In this work, we propose a vehicular channel prediction framework that combines sequential learning and attention-based temporal modeling to capture both short-term dynamics, corresponding to rapid channel fluctuations caused by fast fading and Doppler effects, and long-range temporal evolution induced by changes in vehicle position, propagation geometry, and blockage conditions. The framework is trained and evaluated on the real measured MaMIMOSA dataset \cite{mami_mosa}, ensuring that the predicted channels preserve realistic spatiotemporal structure and mobility effects. Unlike most existing approaches, we further assess prediction quality at the system level by analyzing its impact on achievable SE.
Prior learning-based approaches for vehicular channel prediction include CNN--LSTM architectures for Doppler-aware forecasting \cite{gao_v2v_pred,rnn_v2x}, Transformer-based CSI predictors \cite{transformer_v2x}, and mobility-aware methods incorporating graph or spatiotemporal information \cite{gnn_v2x, spatiotemporal_v2x}. While these approaches demonstrate promising prediction accuracy, most are evaluated using simulated or semi-synthetic channel models and therefore do not fully capture the non-stationary and environment-dependent characteristics of real vehicular propagation.
As highlighted in comprehensive surveys \cite{survey_v2x}, data scarcity remains a fundamental challenge in vehicular communication, significantly limiting the generalization capability of purely data-driven models and motivating the need for realistic, measurement-driven augmentation strategies.
In addition, existing works predominantly assess performance using prediction-oriented metrics such as mean squared error (NMSE), while rarely validating the impact of channel prediction accuracy at the communication system level. In particular, the effect of predicted CSI on achievable SE, a critical performance indicator for 6G vehicular systems, is often not evaluated. This lack of system-level validation leaves an important gap between channel prediction accuracy and its practical benefit for vehicular communication performance. Recent studies have demonstrated that LSTM-based channel predictors can outperform classical channel prediction techniques, such as Wiener filtering and Kalman-based methods, in time-varying wireless environments by better capturing nonlinear temporal channel dynamics \cite{Kibugi2021,Kim2020}. Consequently, LSTM has become a widely adopted learning-based benchmark for wireless channel prediction. In this work, LSTM is therefore used as the baseline model upon which more advanced temporal learning architectures are evaluated. To address the limitations of recurrent learning, we investigate four complementary temporal learning paradigms for vehicular CSI prediction: recurrent learning (LSTM), convolutional temporal modeling (TCN), encoder-based attention (CNN+Transformer), and decoder-only Transformer modeling (ChannelGPT). These architectures were selected because they represent fundamentally different approaches for modeling temporal dependencies, ranging from sequential memory mechanisms to attention-based sequence modeling. To the best of our knowledge, a unified comparison of these temporal learning paradigms using real measured vehicular distributed multiple-
input multiple-output (MIMO) channels has not been previously reported. Such a comparison provides valuable insights into the trade-offs between prediction accuracy, forecasting robustness, and computational complexity under realistic vehicular propagation conditions.

This paper makes the following key contributions:
\begin{itemize}
\item We leverage the real-world MaMIMOSA vehicular measurement campaign to investigate non-stationary and mobility-driven channel behavior in a distributed-MIMO setting using measured CSI data.

\item We develop a unified benchmarking framework comparing recurrent, convolutional, encoder-based, and decoder-only Transformer architectures for vehicular CSI prediction under identical preprocessing, training, and evaluation conditions.

\item We evaluate both single-step and multi-horizon CSI prediction performance and further assess the impact of predicted channels at the system level through spectral-efficiency analysis, establishing the relationship between prediction accuracy and communication performance.

\end{itemize}

The remainder of this paper is organized as follows. Section~II describes the measurement campaign and dataset. Section~III details channel representation and preprocessing. Section~IV presents the proposed learning framework. Section~V outlines the experimental setup and evaluation metrics. Section~VI discusses the prediction and SE results. Section~VII concludes the paper.
\section{Measurement Campaign}

\begin{figure}
    \centering
    \includegraphics[width=0.8\linewidth]{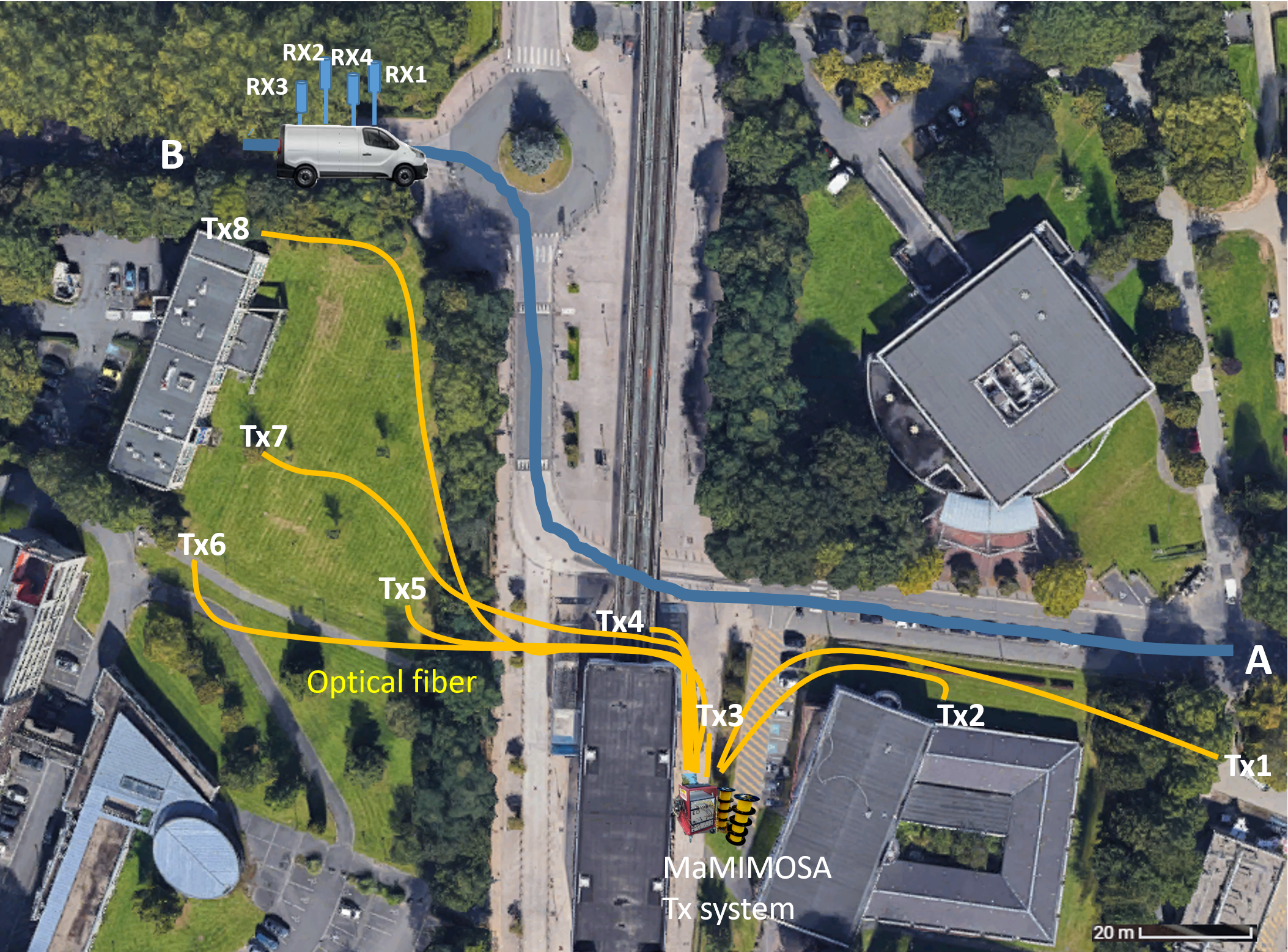}
    \caption{Top view of the measurement campaign showing eight distributed APs and four antennas mounted on the van.
}
    \label{fig:drivingtest}
\end{figure}

The channel measurements were conducted on the ``Cité Scientifique'' campus in Lille, France, using the MaMIMOSA channel sounder. The campaign targets realistic 5.89~GHz vehicular propagation conditions and serves as the experimental basis for this study. As illustrated in Fig.~\ref{fig:drivingtest}, $4$ receiving antennas were mounted on a vehicle moving along a $200$~m campus boulevard at an approximately constant speed of $20$~km/h, ensuring smooth temporal sampling of channel variations. The measurement campaign corresponds to uplink (UL) channel acquisition between the vehicle and the distributed APs. Under the assumption of channel reciprocity in TDD systems, the measured channels are subsequently reused for downlink (DL) system-level performance evaluation, i.e., acquisition from $8$ distributed single-antenna APs. 
% Under the assumption of channel reciprocity in time division duplex (TDD) systems, the measured channels can be used for both DL and uplink (UL) system-level performance evaluation.
%The scenario considers a single mobile user represented by the vehicle equipped with 4 transmit antennas, while the distributed access points (APs), each equipped with an antenna, act as receivers. Each transmit antenna contributes to spatial multiplexing, enabling the transmission of parallel data streams toward the APs, with each AP receiving one symbol per transmission, resulting in $8$ parallel received streams. This configuration emulates a vehicular distributed MIMO system and enables the analysis of spatial diversity and parallel channel observations. 
It is worth mentioning that our campaign-measurement  captures heterogeneous propagation conditions, including line-of-sight (LOS) and obstructed scenarios with scattering and blockage effects. This distributed configuration reflects practical vehicular distributed MIMO deployments and enables the evaluation of spatial variability. In this campaign, the channel was recorded over approximately 52 frames, corresponding to a total measurement time of about 35 seconds, with each frame consisting of 128 symbols (i.e., measurements). This temporal resolution captures both small-scale fading and mobility-induced large-scale variations, making the dataset well suited for validating learning-based channel prediction and generation methods using real vehicular measurements.

\section{Vehicular distributed MIMO Channel Representation and Preprocessing}
\subsection{Channel Data Representation}

The vehicular wireless channel is represented as a five-dimensional 
complex-valued tensor
\begin{equation}
    \mathbf{H} \in \mathbb{C}^{F \times T \times A \times M \times K},
\end{equation}
where $F$ denotes the number of subcarriers, $T$ the number of symbols, $A$ the number of frames, $M$ the number of APs, and $K$ denotes the number of transmit antennas at the vehicle (single-user scenario). This
representation captures the joint frequency, temporal, and spatial
evolution of vehicular distributed-MIMO channels. Each channel coefficient is a complex number expressed as
\begin{equation}
    h = h_{\mathrm{Re}} + j h_{\mathrm{Im}},
\end{equation}
where $h_{\mathrm{Re}}$ and $h_{\mathrm{Im}}$ denote the real and imaginary
components, respectively. This representation preserves both amplitude
and phase information required for coherent detection and channel
prediction. The key characteristics of the dataset are summarized in
Table~\ref{tab:channel_stats}.

\begin{table}[h]
\centering
\caption{Channel Dataset Structure and Key Statistics}
\label{tab:channel_stats}
\begin{tabular}{l c}
\hline
\textbf{Property} & \textbf{Value} \\
\hline
Tensor dimensions $(F,T,A,M,K)$ & $(818,128,52,8,4)$ \\
Data type & Complex-valued \\
Std. dev. (real/imag) & $\approx 8.4 \times 10^{-5}$ \\
\hline
\end{tabular}
\end{table}
\subsection{Channel Data Transformation}

Let $\mathbf{H} \in \mathbb{C}^{F \times T \times A \times M \times K}$ denote the raw channel tensor. 
We denote by $h_{f,m,k}(t) \in \mathbb{C}$ the channel coefficient at subcarrier $f$, time index $t$, AP antenna $m$, and number of user antennas $k$. The $T$ symbols per frame and $A$ frames are concatenated into a single temporal index $t \in \{0, \dots, T \times A - 1\}$, such that $T \times A$ defines the overall temporal dimension of the data. To enable learning with real-valued neural networks, each complex coefficient is decomposed into its real and imaginary components. 
To reduce dimensionality, we select $64$ subcarriers, denoted by $F_{\text{sel}}$. The selection is computed on the training set only to avoid data leakage. For each AP, the selected coefficients are flattened into a real-valued feature vector. The dimension $D$ corresponds to the feature frequential dimension, while $T \times A$ represents the temporal dimension:
\begin{equation}
\mathbf{h}(t) \in \mathbb{R}^{D}, \quad
D = 2 \times F_{\text{sel}} \times M \times K,
\end{equation}
where the factor of $2$ accounts for the concatenation of real and imaginary components. Feature-wise normalization is performed using statistics $\mu$ and $\sigma$ computed exclusively from the training set:
\begin{equation}
\mathbf{h}(t) =
\frac{(\mathbf{h}(t) - \mu)}{\sigma}.
\end{equation}
which stabilizes optimization and improves generalization. Here,
${\mu}$ is the per-feature mean and
$\boldsymbol{\sigma}$ is the per-feature standard deviation. Finally, a sliding window of length $W$ constructs input--target pairs $(\mathbf{X}_t, \mathbf{y}_t)$, where $\mathbf{X}_t$ denotes the input sequence and $\mathbf{y}_t$  the corresponding ground-truth next-step channel:

\begin{align}
    \mathbf{X}_t &=
    \bigl[\mathbf{h}(t - W + 1), \ldots, \mathbf{h}(t)\bigr]
    \in \mathbb{R}^{W \times D}, \\
    \mathbf{y}_t &= \mathbf{h}(t+1) \in \mathbb{R}^{D}.
\end{align}
\section{Proposed Channel Prediction Methodology}

The objective of the proposed framework is to predict realistic channel realizations that preserve the intrinsic spatiotemporal structure of measured vehicular distributed MIMO data. This facilitates the development of robust learning-based communication strategies, including SE optimization and channel prediction. 
% {\color{blue} To this end, we investigate multiple complementary temporal learning architectures for vehicular CSI prediction under time-varying and non-stationary propagation conditions. LSTM is considered as a recurrent baseline for sequential dependency learning, while TCN captures multi-scale temporal correlations through causal convolutions. The hybrid CNN+Transformer combines local feature extraction with global attention to jointly model short-term channel fluctuations and long-range temporal dependencies. In addition, a decoder-only Transformer architecture, referred to as ChannelGPT, is included as an attention-based sequence modeling baseline for realistic vehicular distributed-MIMO CSI prediction.}

\subsection{LSTM-Based Channel Predictor}
\begin{figure}
    \centering
    \includegraphics[width=0.9\linewidth]{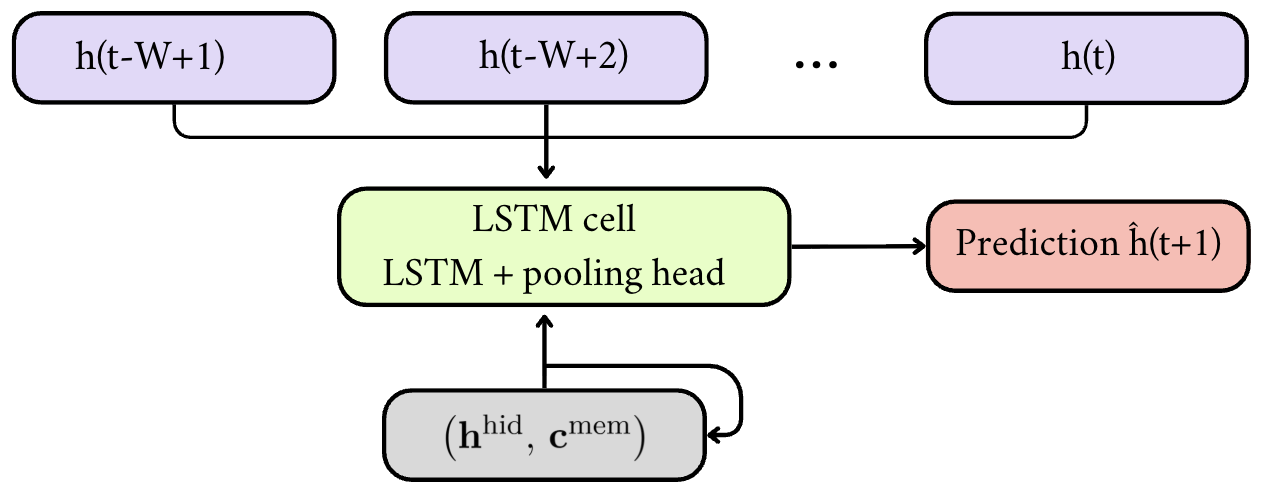}
    \caption{LSTM-based sequence-to-one channel predictor.}
    \label{fig:LSTM}
\end{figure}

Long Short-Term Memory (LSTM) networks mitigate vanishing-gradient effects in recurrent models by maintaining gated hidden and memory states \cite{hochreiter1997}.The LSTM maintains a hidden state $\mathbf{h}^{\text{hid}}$ and a memory cell $\mathbf{c}^{\text{mem}}$ to capture temporal dependencies. The memory cell stores information from previous channel observations, while the hidden state summarizes the relevant temporal context required for prediction, enabling the model to capture channel evolution over time. In our framework, the LSTM predictor follows a sequence-to-one formulation using a sliding window of $W$ past channel vectors $\{\mathbf{h}(t-W+1), \dots, \mathbf{h}(t)\}$, and outputs the next-step estimate $\hat{\mathbf{h}}(t+1)$, as illustrated in Fig.~\ref{fig:LSTM}. Stacked unidirectional LSTM layers process the input sequence, and the resulting hidden states are aggregated via hybrid pooling (mean, max, and attention). A regression head maps the pooled representation to the predicted channel vector. Channel sequences are constructed independently per AP to avoid information leakage, with complex coefficients represented using real and imaginary components. The loss function combines Smooth~L1 with a cosine similarity term to promote robustness and directional consistency.

\subsection{TCN-Based Channel Predictor}
\begin{figure}
    \centering
    \includegraphics[width=1.0\linewidth]{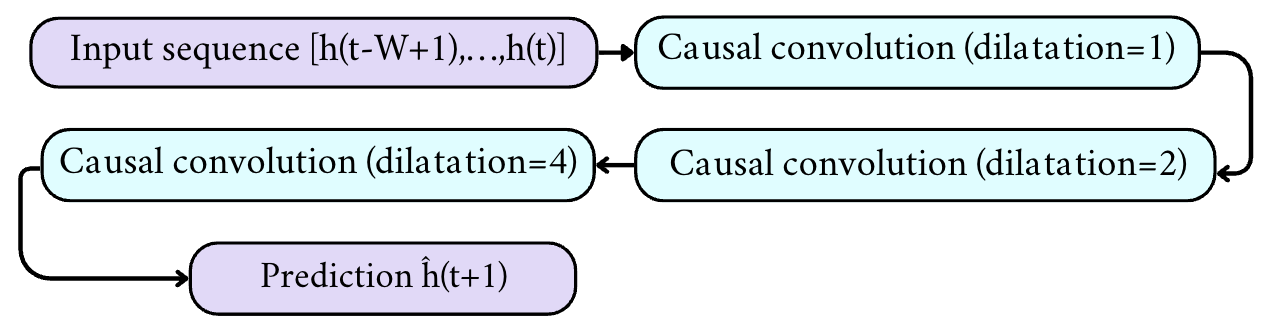} 
    \caption{TCN-based next-step channel predictor with causal dilated residual blocks.}
    \label{fig:TCN}
\end{figure}

Temporal Convolutional Networks (TCNs) provide a non-recurrent alternative for sequence modeling using causal dilated convolutions \cite{bai2018}. The proposed TCN processes an input window of length $W$ through a stack of residual convolutional blocks with increasing dilation (e.g., 1, 2, 4), ensuring a large receptive field while preserving causality, as illustrated in Fig.~\ref{fig:TCN}. A last-timestep readout summarizes the temporal context and is mapped to the next-step channel prediction. Channel sequences are constructed independently per AP to avoid cross-AP leakage. Residual connections, normalization, and gated activations improve training stability, while Smooth~L1 regression ensures robust convergence.

\subsection{Hybrid CNN~+~Transformer Encoder Channel Predictor}
\begin{figure}
    \includegraphics[width=1.0\linewidth]{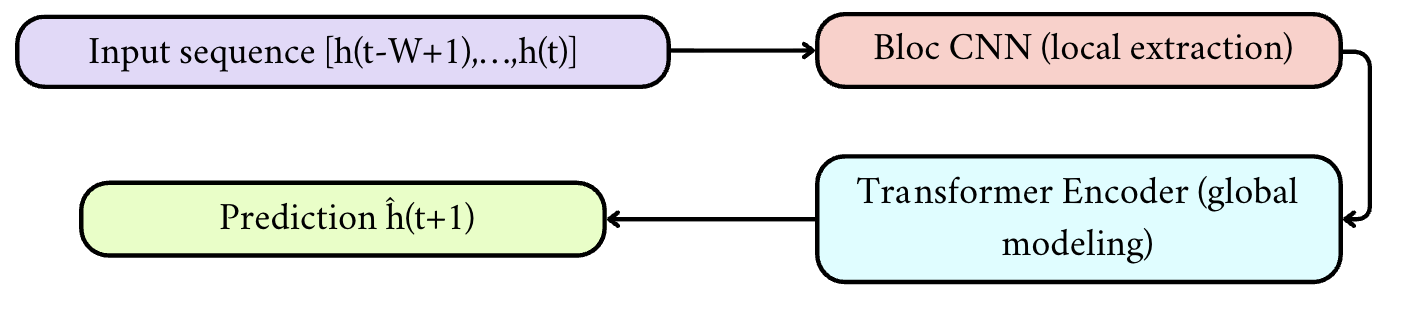}
    \caption{Hybrid CNN~+~Transformer encoder for next-step vehicular distributed MIMO channel prediction.}
    \label{fig:cnn_transformer}
\end{figure}

To capture both local temporal patterns and long-range dependencies, we propose a hybrid CNN~+~Transformer encoder architecture, illustrated in Fig.~\ref{fig:cnn_transformer}. A temporal CNN stem with residual depthwise-separable convolutions extracts short-term dynamics and projects features into a shared embedding space. The resulting sequence is processed by pre-norm Transformer encoder blocks with multi-head self-attention and positional encodings (with optional ALiBi bias) to model temporal dependencies. The encoded representation is summarized using last-step or last-$k$ pooling and mapped to the next-step channel estimate $\hat{\mathbf{h}}(t+1)$.

\subsection{ChannelGPT-Based Channel Generator}

\begin{figure}
    \includegraphics[width=1.0\linewidth]{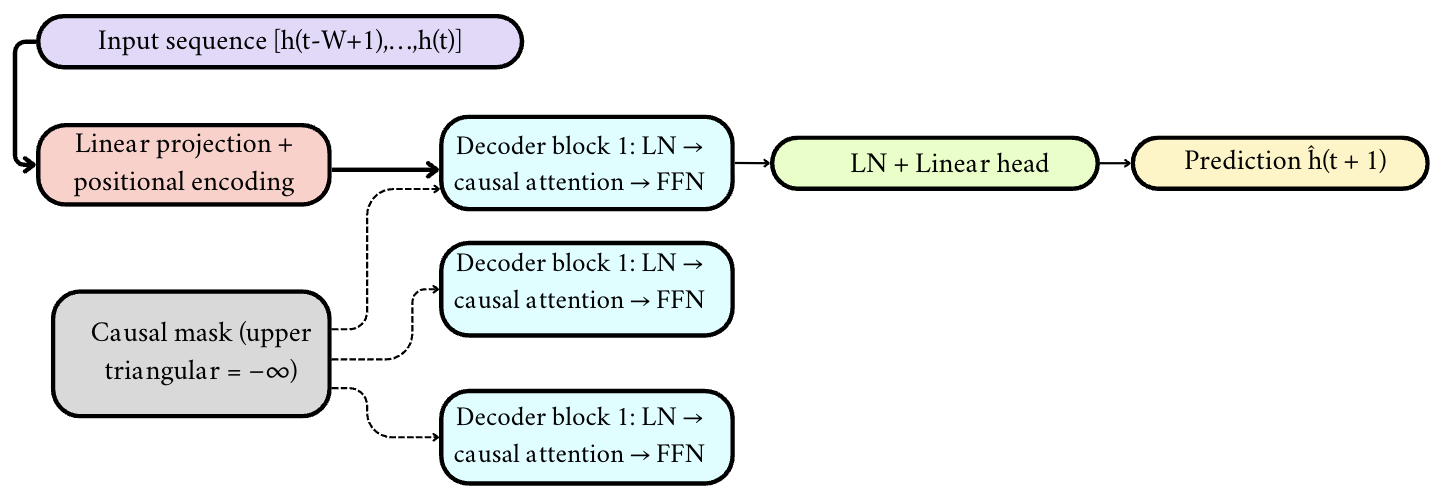}
    \caption{ChannelGPT-based autoregressive channel sequence generator.}
    \label{fig:channelgpt}
\end{figure}

We adopt ChannelGPT, a decoder-only Transformer for generative CSI modeling, as illustrated in Fig.~\ref{fig:channelgpt}. The model first applies a linear projection with positional encoding to the input sequence, followed by a stack of Transformer blocks composed of layer normalization (LN), multi-head self-attention, and feedforward networks (FFN). The final hidden representation is passed through a linear regression head that directly predicts future channel realizations over a predefined forecasting horizon. Specifically, given an observed window of length $W$, ChannelGPT predicts the next channel realization $\hat{\mathbf{h}}(t+1)$. Prediction performance is subsequently evaluated across multiple forecasting horizons, as discussed in Section~VI-B. The model operates on real-valued vectors obtained from unfolded complex CSI, using leakage-free preprocessing.

\section{Performance evaluation}
\subsection{Evaluation of predictions}
To evaluate the effectiveness of the proposed prediction framework, a series of controlled experiments are conducted under realistic vehicular distributed MIMO channel prediction settings. Channel sequences are preprocessed as described in Section~II and subsequently partitioned into training, validation, and test sets to ensure statistically sound evaluation. The dataset was partitioned chronologically into training (80\%), validation (10\%), and test (10\%) sets to avoid temporal leakage between highly correlated samples. Normalization statistics and subcarrier selection were computed exclusively on the training set. Although the dataset was collected along a single measurement route, it exhibits substantial spatial and frequency diversity. In the spatial domain, the distributed-MIMO deployment consists of eight geographically distributed access points communicating with four vehicle-mounted antennas, resulting in multiple independent propagation links. In the frequency domain, the channel is observed over 64 selected OFDM subcarriers, capturing frequency-selective fading and rich multipath propagation effects. Furthermore, both LOS and obstructed propagation conditions are present throughout the measurement campaign, providing a realistic benchmark for measurement-driven vehicular CSI prediction.
The primary evaluation metric is the NMSE, which quantifies the reconstruction error relative to the energy of the true channel realization:
\begin{equation}
    \mathrm{NMSE} = \frac{\| \mathbf{y} - \widehat{\mathbf{y}} \|_2^2}{\| \mathbf{y} \|_2^2},
\end{equation}
where $\mathbf{y}$ and $\widehat{\mathbf{y}}$ denote the true and predicted channel vectors, respectively. NMSE provides a scale-invariant measure of prediction accuracy and is particularly well-suited for wireless channel modeling, where channel magnitudes can vary significantly across propagation conditions.

\subsection{Model Parameters}
All models were trained under identical optimization settings to ensure a fair comparison. A batch size of 128, an input window of length $W = 48$, and 200 training epochs were used for all architectures. Optimization was performed using AdamW with a learning rate of $10^{-3}$ and weight decay of $10^{-5}$. Gradient clipping (max norm 1.0) and a OneCycle learning rate schedule were applied to stabilize training. This unified protocol ensures that performance differences primarily reflect architectural design rather than optimization choices (Table~\ref{tab:Model_params}). The model configurations balance capacity and efficiency for capturing temporal dynamics, with recurrent and convolutional models handling multi-scale dependencies, and attention-based models enabling richer temporal representations. ChannelGPT employs a decoder-only Transformer architecture for multi-step vehicular CSI prediction.

\begin{table*}[!t]
\centering
\caption{Unified Architecture and Hyperparameter Summary for Sequential and Generative Models}
\label{tab:Model_params}
\begin{tabular}{l
                c
                c
                c
                p{3.8cm}
                p{1.6cm}
                p{2.4cm}}
\toprule
\textbf{Model} & \textbf{Layers} & \textbf{Hidden Dim.} & \textbf{FF Dim.} &
\textbf{Core Mechanism / Key Config.} &
\textbf{Att./Kernel / Heads} &
\textbf{Special Options} \\
\midrule

LSTM & 2 & 128 & 128 &
Input proj.\ $\to$128, Huber + cosine loss
& --
& Dropout 0.5, noise $\sigma=0.05$
\\

TCN & 5 & 384 & -- &
Proj.\ $D\!\to\!192$, 5 dilated causal GLU blocks, last-timestep head
& $k=5$
& OneCycleLR, Huber + cosine loss, bf16 AMP
\\

CNN + Transformer Encoder & 3 CNN + 6 Enc. & 256 & 1536 &
CNN stem + self-attention (ALiBi, SwiGLU)
& $k=7$, 8 heads
& --
\\

ChannelGPT (Decoder-only) & 6 & 256 & 1024 &
Decoder-only
Transformer (GPT-2 style,
learned PE)
& 8 heads& --

\\

\bottomrule
\end{tabular}

\vspace{1mm}
\footnotesize{\textit{Note:} FF = Feedforward, Attn. = Attention, FC = Fully Connected, PE = Positional Encoding.}
\end{table*}

\vspace{1mm}

\subsection{System-level performance evaluation}
We consider a single-user vehicular-to-infrastructure (V2I) UL communication system where a vehicle (user equipment, UE) equipped with $K$ transmit antennas communicates with $M$ geographically distributed single-antenna APs. The vehicle simultaneously transmits $M$ independent data streams, with each AP responsible for decoding a single stream. Let $\mathbf{s}_f \in \mathbb{C}^{M \times 1}$ denote the vector of transmitted symbols at subcarrier $f$, where the elements $s_{f,m} \sim \mathcal{CN}(0,1)$ are independent and identically distributed (i.i.d.) complex Gaussian symbols. To control the transmit power across streams and subcarriers, we introduce a per-stream power allocation strategy ~\cite{interdonato2019local_pzf}. Specifically, the transmit signal at subcarrier $f$ is expressed as
\begin{equation}
\mathbf{x}_f = \mathbf{W}_f \mathbf{P}_f^{1/2} \mathbf{s}_f,
\end{equation}
where $\mathbf{W}_f \in \mathbb{C}^{K \times M}$ denotes the precoding matrix, $\mathbf{s}_f \in \mathbb{C}^{M \times 1}$ is the vector of normalized data symbols satisfying $\mathbb{E}[\mathbf{s}_f \mathbf{s}_f^H] = \mathbf{I}_M$, and $\mathbf{P}_f = \mathrm{diag}(p_{1,f}, \dots, p_{M,f}) \in \mathbb{R}^{M \times M}$ is a diagonal matrix containing the power allocated to each stream. To enable low-complexity reception at the APs, we adopt a use-and-then-forget (UatF) framework~\cite{doan2021massive_keyhole_uatf} combined with PZF precoding~\cite{interdonato2019local_pzf} at the transmitter. The key idea is that all signal processing is performed at the vehicle, such that each AP can detect its intended stream without requiring additional processing or instantaneous CSI. This enables us to quantify the impact of prediction accuracy on system performance. Table~\ref{tab:eval_params} summarizes the main parameters used in the evaluation.
\begin{table}
\centering
\caption{Parameters Used for Channel Prediction and SE Evaluation}
\label{tab:eval_params}
\begin{tabular}{|c|l|c|}
\hline

\textbf{Symbol} & \textbf{Description} & \textbf{Value} \\

\hline
\( w_p \) & Noise level (dBm) & $-93$ dBm \\
\hline
\( \sigma^2 \) & Noise power & $0.001 \times 10^{w_p/10}$ \\
\hline
\( P_{\text{tx}} \) & Total transmit power & 1 W \\
\hline

\( \tau_p \) & Pilot length & 4 \\
\hline
\( \tau_c \) & Coherence block length & 168 \\
\hline
\( \xi \) & Fraction of DL data symbols & 0.5 \\
\hline
% \( \eta \) & DL overhead factor & $\xi(1 - \tau_p/\tau_c)$ \\
% \hline

% \( \mathbf{w}_k \) & Precoding scheme used & ZF (Zero-Forcing) \\
% \hline

\end{tabular}
\end{table}

In addition, the DL signal-to-interference-plus-noise ratio (SINR) and the corresponding SE are computed for each user under PZF precoding. In a distributed setup with $M$ APs, the received signal of user antenna $k$ at channel use $n$ is the coherent superposition of the contributions transmitted by all APs. Following the UatF bounding technique, the effective SINR of user antenna $k$, given by \eqref{eq:sinr_like_figure}
\begin{equation}
\mathrm{SINR}_{m,f}
=
\frac{\left| CP_{m,f} \right|^{2}}
{\mathbb{E}\!\left[\left|PGU_{m,f}\right|^{2}\right]
+\sum\limits_{m'\neq m}\mathbb{E}\!\left[\left|MSI_{m',f}\right|^{2}\right]
+\sigma_n^2},
\label{eq:sinr_like_figure}
\end{equation}
where $\sigma_n^2$ is the receiver noise power.
We define the following terms:

\begin{align}
CP_{m,f}
&\triangleq
\mathbb{E}\!\left[
\sqrt{p_{m,f}}\,
\mathbf{h}_{m,f}^{H}\mathbf{w}_{m,f}
\right],
\label{eq:cp_def}
\\[2mm]
PGU_{m,f}
&\triangleq
\sqrt{p_{m,f}}\,
\mathbf{h}_{m,f}^{H}\mathbf{w}_{m,f}
-
CP_{m,f},
\label{eq:pgu_def}
\\[2mm]
MSI_{m^\prime,f}
&\triangleq
\sqrt{p_{m^\prime},f}\,
\mathbf{h}_{m,f}^{H}\mathbf{w}_{m^\prime,f},
\qquad m^\prime\neq m,
\label{eq:mui_def}
\end{align}

In \eqref{eq:cp_def}, $\mathbf{h}^{\mathrm{H}}_{m,f} \in \mathbb{C}^{1\times K}$ denotes the DL channel coefficients between AP $m$ and vehicle antennas at frequency $f$. $\mathbf{w}_{m,f} \in \mathbb{C}^{K \times 1}$ is the precoding vector at frequency $f$. Hence, $CP_{m,f}$ represents the coherent mean desired signal, $PGU_{m,f}$ captures the random fluctuation around this mean (precoding gain uncertainty), $MSI_{m^\prime,f}$ is the multi-stream interference from AP $m^\prime$.
The achievable DL SE of subcarrier $f$ is given by
\begin{equation}
\mathrm{SE}_f
=
\sum_{m=1}^{M}
\underbrace{\xi \left( 1 - \frac{\tau_p}{\tau_c} \right)}_{\text{prelog factor}}
\log_2 \!\Bigl( 1 + \mathrm{SINR}_{m,f} \Bigr),
\label{eq:se_pzf}
\end{equation}
where the prelog factor accounts for the fraction of symbols used for UL data transmission after pilot overhead. When channel prediction is employed and no pilot-based channel estimation is required, we set $\tau_p = 0$, resulting in the prelog factor $\xi$. 
% For conventional channel estimation, the pilot length is $\tau_p = 8$, which yields the prelog factor
% \begin{equation}
% \xi\left(1 - \frac{\tau_p}{\tau_c}\right).
% \end{equation}

% This formulation enables a direct comparison of the achievable SE obtained using estimated channels and predicted or generated channels, while explicitly accounting for pilot overhead.

\begin{table}[t]
\centering
\caption{Comparison of validation and test NMSE (lower is better).}
\label{tab:comparison_nmse}
\begin{tabular}{lcccc}
\toprule
\textbf{Model} & \textbf{Val NMSE} & \textbf{Ep.} & \textbf{Test NMSE} & \textbf{Ep.} \\
\midrule
CNN + Transformer & 0.0221 & 161 & \textbf{0.0290} & 161 \\
ChannelGPT     & \textbf{0.0215} & 118 & 0.0331 & 176 \\
TCN            & 0.0269 & 175 & 0.0332 & 162 \\
LSTM           & 0.3751 & 165 & 0.6949 & 189 \\
\bottomrule
\end{tabular}
\end{table}

\section{Numerical Results}
\subsection{Single-Step Prediction Performance}

\begin{figure}
    \centering
    \includegraphics[width=0.8\linewidth]{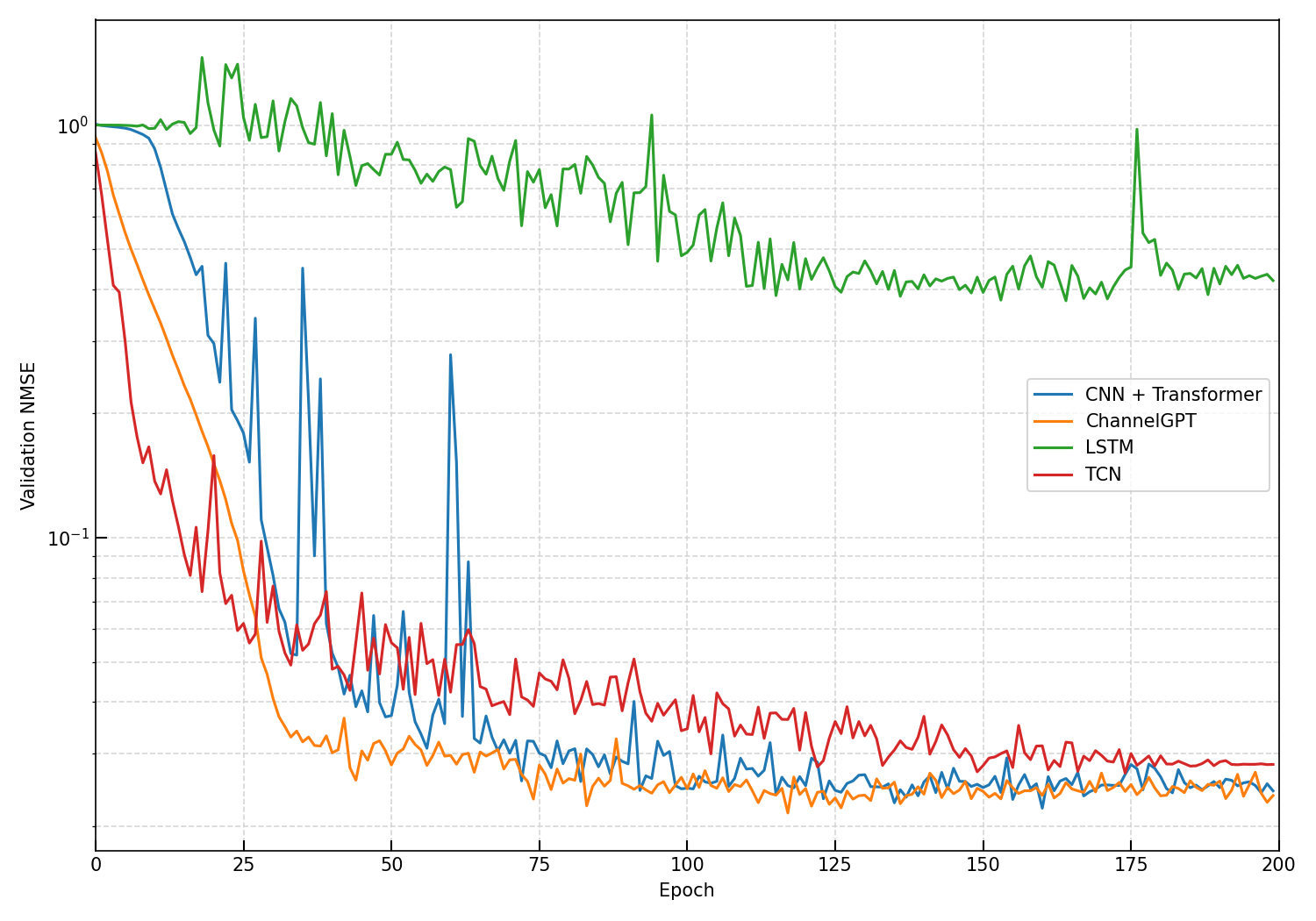}
    \caption{Validation NMSE comparison among LSTM, TCN, CNN~+~Transformer, and ChannelGPT.}
    \label{fig:val_curves}
\end{figure}

We first compare the single-step CSI prediction performance of the considered models in terms of convergence behavior, validation accuracy, and test-set generalization. Fig.~\ref{fig:val_curves} compares the training and validation NMSE over 200~epochs for all models. The LSTM baseline converges slowly and stabilizes at a high error level, reflecting limited ability to capture long-range temporal dependencies. In contrast, TCN and CNN~+~Transformer converge rapidly within the first 50~epochs and exhibit stable training behavior. ChannelGPT achieves the most consistent convergence, combining fast error decay with minimal oscillations, indicating robust modeling of nonlinear temporal dynamics in vehicular distributed MIMO  channels.
Quantitative results in Table~\ref{tab:comparison_nmse} confirm these observations. ChannelGPT and CNN~+~Transformer achieve over 94\% NMSE reduction compared to LSTM, while TCN also provides a significant improvement (92.8\%). ChannelGPT reaches its optimal validation performance earlier than the other models, highlighting faster and more stable convergence.
Test results further demonstrate superior generalization of attention-based models. CNN~+~Transformer achieves the lowest test NMSE (0.0290), followed closely by ChannelGPT (0.0331) and TCN (0.0332), whereas LSTM exhibits a substantially higher error (0.6949). Overall, attention-centered architectures, particularly CNN~+~Transformer and ChannelGPT, deliver the best accuracy and robustness, achieving an order-of-magnitude improvement over recurrent baselines for vehicular distributed MIMO channel prediction.

\subsection{Multi-Horizon Prediction Performance}

\begin{figure}
    \centering
    \includegraphics[width=0.8\linewidth]{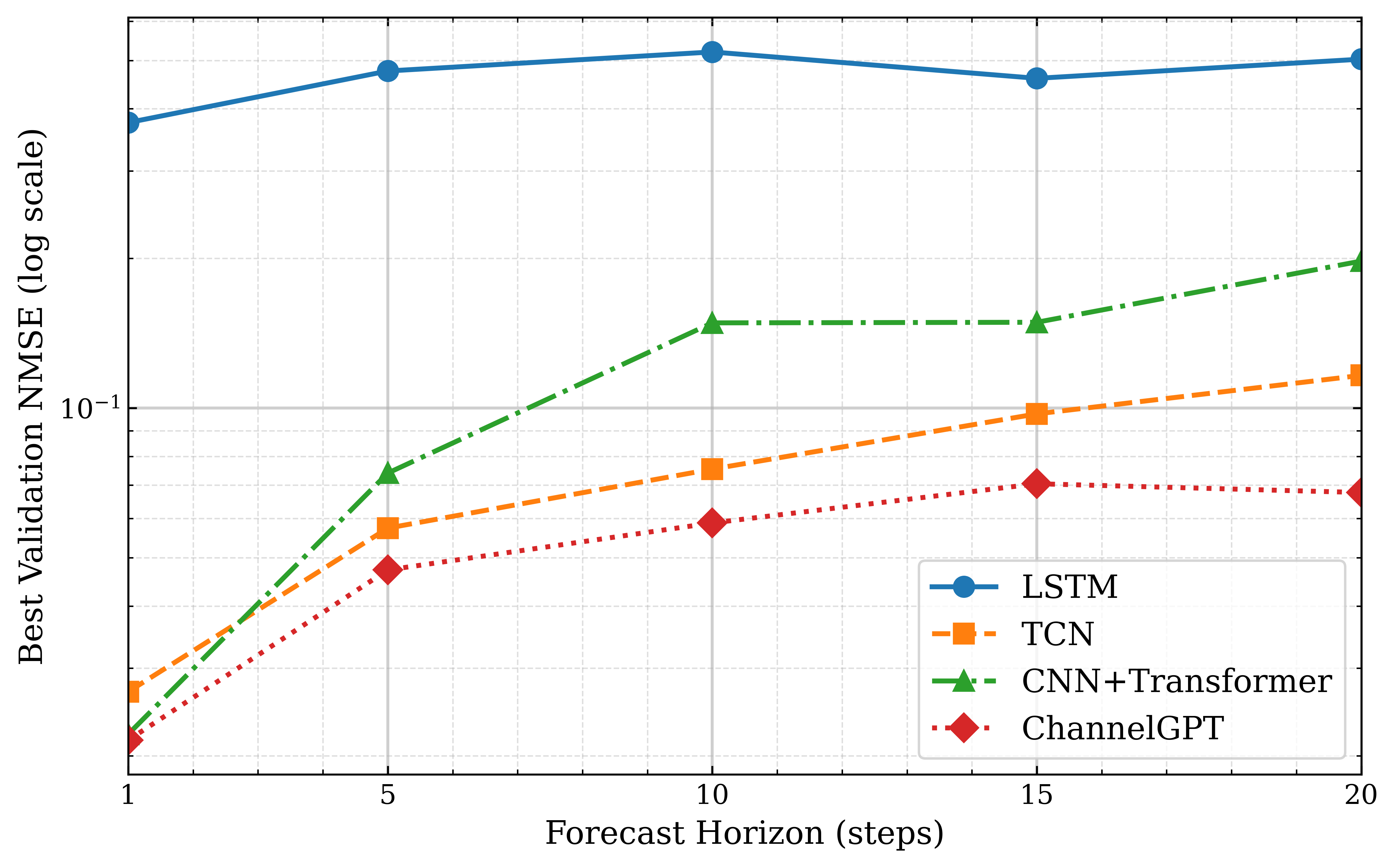}
    \caption{NMSE as a function of prediction horizon for all considered models.}
    \label{fig:horizon_nmse}
\end{figure}
To assess robustness beyond one-step forecasting, the trained models are evaluated at multiple prediction horizons. Specifically, prediction accuracy is reported for horizons $T_{\mathrm{pred}} \in \{1,5,10,15,20\}$, corresponding to increasingly distant future channel realizations. This evaluation provides insight into each model's ability to preserve temporal channel dynamics as the forecasting interval increases.
Fig.~\ref{fig:horizon_nmse} presents the prediction NMSE as a function of the forecasting horizon. As expected, prediction accuracy degrades as the horizon increases due to the growing uncertainty in future channel evolution. While CNN+Transformer achieves the lowest single-step test NMSE, ChannelGPT maintains the lowest NMSE across longer prediction horizons, demonstrating improved robustness for long-term vehicular CSI forecasting. The results indicate that attention-based architectures are more effective than recurrent approaches at preserving temporal channel dynamics over extended forecasting intervals. The observed trends can be explained by the different temporal modeling capabilities of the considered architectures. The LSTM baseline exhibits the highest NMSE because recurrent models compress historical observations into hidden states, which may limit their ability to preserve temporal information over extended time intervals under non-stationary vehicular propagation conditions. TCN improves prediction accuracy through causal dilated convolutions that capture multi-scale temporal correlations more effectively. CNN+Transformer achieves the lowest single-step test NMSE by combining local temporal feature extraction with global self-attention, enabling the modeling of both short-term channel fluctuations and longer-range temporal evolution. In contrast, ChannelGPT maintains the lowest NMSE across longer forecasting horizons while requiring fewer parameters, lower computational complexity, and shorter inference time, resulting in a favorable accuracy--complexity trade-off for long-term vehicular CSI prediction.
\subsection{System-Level Performance Validation}

Beyond system-level accuracy, it is essential to verify whether improved channel prediction translates into tangible system-level gains. 
\begin{figure} \centering \includegraphics[width=0.8\linewidth]{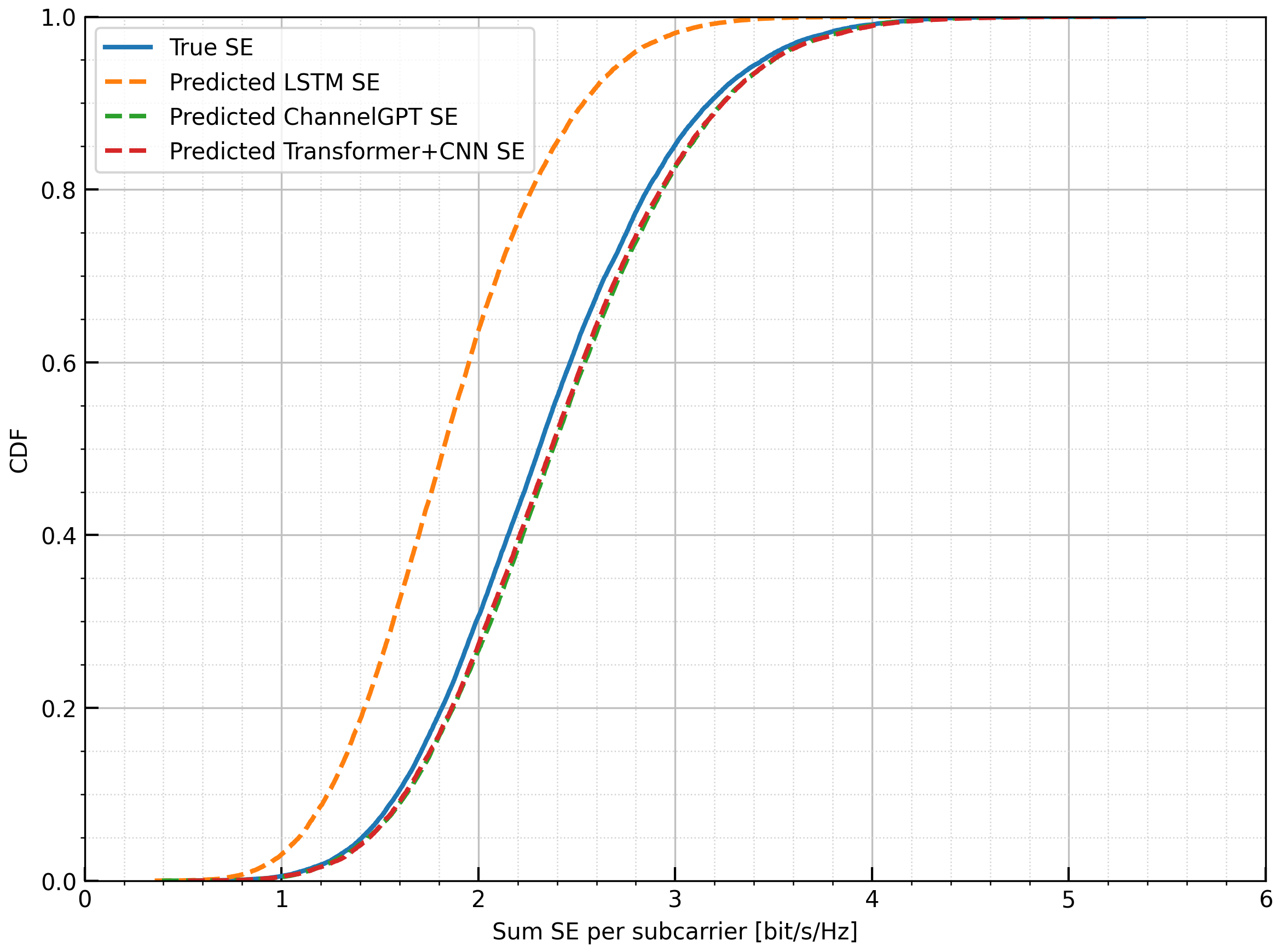} \caption{Cumulative distribution function (Cdf) of SE comparison of different models using the channel estimation} \label{fig:cdf_SE_results} \end{figure}
Figure~\ref{fig:cdf_SE_results} shows that the predictions obtained with ChannelGPT and CNN~+~Transformer closely match the true SE distribution. For instance, at the median point (CDF $\approx$ 0.5), the true SE is around 2.2 bit/s/Hz, while ChannelGPT and CNN~+~Transformer provide very similar values (about 2.15--2.25 bit/s/Hz). Likewise, at higher percentiles (CDF $\approx$ 0.9), the true SE reaches approximately 3.4 bit/s/Hz, and both models remain very close to this value. In contrast, the LSTM model shows a noticeable deviation from the true distribution, with a predicted SE of about 1.8 bit/s/Hz at CDF $\approx$ 0.5, indicating an underestimation of the SE. Overall, these results demonstrate that ChannelGPT and CNN~+~Transformer provide predictions that are much closer to the true SE distribution, achieving better performance than the LSTM model.
% \begin{figure}
%     \centering
%     \includegraphics[width=0.8\linewidth,height=0.6\linewidth]{images/cdf_perfect_csi_vs_pzf.png}
%     \caption{CDF of sum SE per subcarrier using predicted CSI and pilot-based channel estimation (PZF). {to be modified}}
%     \label{fig:cdf_prediction_vs_estimation}
% \end{figure}

% Fig.~\ref{fig:cdf_prediction_vs_estimation} illustrates the cumulative distribution function (CDF) of the sum SE per subcarrier achieved using predicted CSI and channel estimation (PZF). The predicted CSI consistently outperforms the pilot-based channel estimation across the entire CDF. In particular, the median SE achieved with predicted CSI is approximately $2.8$~bit/s/Hz, whereas channel estimation attains a lower median of around $0.6$~bit/s/Hz. 
% The leftward shift of the PZF curve highlights the impact of estimation errors, noise, and pilot contamination, which degrade the quality of the estimated channel and consequently reduce the achievable SE. In contrast, the predicted CSI yields a right-shifted CDF with higher SE values across most percentiles, demonstrating its ability to more accurately capture channel variations and mitigate estimation impairments.
% These results confirm that channel prediction not only improves channel accuracy but also translates into meaningful system-level gains, yielding higher and more reliable spectral efficiency compared to conventional pilot-based channel estimation in vehicular distributed MIMO  systems.

\section{Model Selection and Performance Analysis}
\begin{table}[t]
\centering
\caption{Accuracy--complexity comparison of the strongest-performing CSI predictors.}
\label{tab:test_complexity}
\setlength{\tabcolsep}{4pt}
\begin{tabular}{lccccc}
\toprule
\textbf{Model} &
\textbf{\begin{tabular}[c]{@{}c@{}}Val.\\NMSE\end{tabular}} &
\textbf{\begin{tabular}[c]{@{}c@{}}Test\\NMSE\end{tabular}} &
\textbf{\begin{tabular}[c]{@{}c@{}}Params\\(M)\end{tabular}} &
\textbf{GFLOPs} &
\textbf{\begin{tabular}[c]{@{}c@{}}Inference\\Time (ms)\end{tabular}} \\
\midrule
TCN & 0.0269 & 0.0332 & 11.37 & 1.074 & 9.45 \\
CNN+Transformer & 0.0221 & \textbf{0.0290} & 9.15 & 0.88 & 3.03 \\
ChannelGPT & \textbf{0.0215} & 0.0331 & \textbf{8.61} & \textbf{0.63} & \textbf{1.86} \\
\bottomrule
\end{tabular}
\end{table}
Beyond prediction accuracy, we compare the prediction accuracy and computational complexity of the strongest-performing architectures. As summarized in Table~\ref{tab:test_complexity}, CNN+Transformer achieves the lowest single-step test NMSE (0.0290), while ChannelGPT attains the lowest validation NMSE and requires the fewest parameters, lowest GFLOPs, and shortest inference time. Although TCN achieves a test NMSE comparable to ChannelGPT, it incurs substantially higher computational complexity. Combined with its strong multi-horizon prediction performance and the SE CDF that remains almost indistinguishable from the perfect-CSI curve, ChannelGPT provides a favorable accuracy--complexity trade-off for real-time vehicular CSI prediction in distributed-MIMO systems. Consequently, ChannelGPT emerges as a promising candidate for vehicular CSI prediction and for reducing reliance on frequent CSI acquisition in distributed-MIMO deployments, while CNN+Transformer remains highly competitive in terms of single-step prediction accuracy.
\section{Conclusion}

This paper presented a data-driven vehicular channel prediction framework based on real urban vehicular distributed-MIMO measurements, benchmarking sequential and Transformer-based temporal learning models including LSTM, TCN, CNN+Transformer, and ChannelGPT. Results demonstrate that attention-based architectures significantly outperform recurrent baselines, achieving more than 94\%  NMSE reduction compared with LSTM. While CNN+Transformer achieves the lowest single-step test NMSE, ChannelGPT provides competitive prediction accuracy together with lower computational complexity and stronger robustness across longer prediction horizons, resulting in a favorable accuracy--complexity trade-off. System-level evaluation confirms that improved prediction accuracy translates into improved spectral efficiency. In particular, the predicted channels obtained with ChannelGPT and CNN+Transformer closely reproduce the spectral-efficiency distribution achieved with true CSI, demonstrating their ability to preserve the dominant temporal and spatial channel characteristics required for distributed-MIMO transmission. These findings highlight the effectiveness of attention-based temporal modeling for accurate vehicular CSI prediction and reliable system-level performance under realistic measurement conditions, supporting future 6G vehicular communications. 
Future work will investigate larger-scale datasets and broader mobility scenarios. Although the MaMIMOSA campaign provides realistic vehicular propagation conditions, evaluating the proposed framework across multiple routes and propagation environments would further assess model generalization.

\section*{Acknowledgment}

This work was supported by the French ANR under the France 2030 RIS3 project (ANR-23-CMAS-0023) and by the RITMEA project, co-funded by the ERDF, the French State, and the Hauts-de-France Regional Council.

% Additional temporal learning architectures, including CNN+LSTM hybrids, graph-based models, and advanced attention-based predictors, will also be explored for multi-step vehicular CSI prediction and system-level performance optimization. Future work will also investigate comparisons with classical statistical channel prediction techniques, including autoregressive (AR/ARMA) models, Kalman filtering, and Doppler-based linear prediction methods, in order to further assess the advantages and limitations of learning-based temporal models under realistic vehicular propagation conditions.

% \section*{Acknowledgment}
% This work builds upon simulated and measured vehicular distributed MIMO  datasets and leverages recent advances in deep learning for temporal channel modeling and generation. The authors acknowledge the contributions of the research community in developing open-source vehicular distributed MIMO  models and channel measurement datasets, which provided the foundation for the experimental validation of this study.

\bibliographystyle{IEEEtran}
\bibliography{references}

% \bibitem{bjornson2017}
% T. Van Chien, E. Björnson, and E. Larsson, ``Cell-free massive MIMO: Uniformly great service for everyone,'' \textit{Proc. IEEE ICC}, 2017.

% \bibitem{vaswani2017}
% A. Vaswani et al., ``Attention is all you need,'' \textit{Proc. NeurIPS}, 2017.

% \bibitem{hochreiter1997}
% S. Hochreiter and J. Schmidhuber, ``Long short-term memory,'' \textit{Neural Computation}, 1997.

% \bibitem{brownlee2021}
% J. Brownlee, ``Deep Learning for Time Series Forecasting,'' Machine Learning Mastery, 2021.

% \bibitem{hochreiter1997}
% S. Hochreiter and J. Schmidhuber,
% ``Long short-term memory,''
% \textit{Neural Computation}, vol. 9, no. 8, pp. 1735–1780, 1997.

% \bibitem{schuster1997}
% M. Schuster and K. K. Paliwal,
% ``Bidirectional recurrent neural networks,''
% \textit{IEEE Transactions on Signal Processing}, vol. 45, no. 11, pp. 2673–2681, 1997.

% \bibitem{bai2018}
% S. Bai, J. Z. Kolter, and V. Koltun,
% ``An empirical evaluation of generic convolutional and recurrent networks for sequence modeling,''
% \textit{arXiv preprint arXiv:1803.01271}, 2018.

% \bibitem{vaswani2017}
% A. Vaswani \textit{et al.},
% ``Attention is all you need,''
% \textit{Advances in Neural Information Processing Systems (NeurIPS)}, 2017.

% \bibitem{krizhevsky2012}
% A. Krizhevsky, I. Sutskever, and G. E. Hinton,
% ``ImageNet classification with deep convolutional neural networks,''
% \textit{Advances in Neural Information Processing Systems (NeurIPS)}, 2012.

% \end{thebibliography}
\end{document}